\RequirePackage{cmap}

\documentclass[usenatbib]{mnras}

\usepackage[T1]{fontenc}
\usepackage[dvipsnames, x11names, svgnames]{xcolor}

\renewcommand{\baselinestretch}{1.0} % Change to 1.65 for double spacing
\usepackage{showyourwork}
\usepackage{amsmath}
\usepackage{amsfonts}
\usepackage{amssymb}
\usepackage{mathrsfs}

\usepackage{xurl}
\usepackage{datetime2}

\usepackage{booktabs}
\usepackage{graphicx}
\usepackage[
]{hyperref}

\usepackage{newtxtext}
\usepackage{newtxmath}

\usepackage{microtype}
\usepackage{csquotes}
\usepackage[capitalise]{cleveref}
\usepackage{xspace}

\usepackage{standalone}
\usepackage{tikz}
\usetikzlibrary{
    calc,
    positioning,
    shapes,
    shapes.symbols,
}

\usepackage{siunitx}
\DeclareSIUnit\parsec{pc}
\DeclareSIUnit\au{au}
\DeclareSIUnit\mas{mas}

\usepackage{aas_macros}
\newcommand{\IWA}{\ensuremath{\mathrm{IWA}}\xspace}
\newcommand{\OWA}{\ensuremath{\mathrm{OWA}}\xspace}
\newcommand{\HWO}{HabWorlds\xspace}
\newcommand{\Decadal}{Astro2020}  % no numbers in macro names

\definecolor{C0}{HTML}{ff0000}
\definecolor{C1}{HTML}{ff6d38}
\definecolor{C2}{HTML}{ecc86f}
\definecolor{C3}{HTML}{a4f89f}
\definecolor{C4}{HTML}{5af8c8}
\definecolor{C5}{HTML}{12c8e6}
\definecolor{C6}{HTML}{386df9}
\definecolor{C7}{HTML}{8000ff}

\newcommand{\rainbows}{
    \textcolor{C0}{r}%
    \textcolor{C1}{a}%
    \textcolor{C2}{i}%
    \textcolor{C3}{n}%
    \textcolor{C4}{b}%
    \textcolor{C5}{o}%
    \textcolor{C6}{w}%
    \textcolor{C7}{s}%
    \xspace
}

\makeatletter
\newcommand{\affdest}[1]{\Hy@raisedlink{\hypertarget{#1}{}}$^\mathrm{#1}$}
\makeatother
\newcommand{\afflink}[1]{\hyperlink{#1}{#1}}

\title{Chasing \rainbows and ocean glints:\\ Inner working angle constraints for the Habitable Worlds Observatory}

\author[Sophia R. Vaughan et al.]{%
    Sophia R. Vaughan\thanks{Correspondence:~\url{sophia.vaughan@physics.ox.ac.uk}}\textsuperscript{,\afflink{1}},
Timothy D. Gebhard\textsuperscript{\afflink{2},\afflink{3}},
Kimberly Bott\textsuperscript{\afflink{4},\afflink{5},\afflink{6}},
Sarah L. Casewell\textsuperscript{\afflink{7}},
\newauthor
Nicolas B. Cowan\textsuperscript{\afflink{8}},
David S. Doelman\textsuperscript{\afflink{9},\afflink{10}},
Matthew Kenworthy\textsuperscript{\afflink{9}},
Johan Mazoyer\textsuperscript{\afflink{11}},
\newauthor
Maxwell A. Millar-Blanchaer\textsuperscript{\afflink{12}},
Victor J. H. Trees\textsuperscript{\afflink{13},\afflink{14}},
Daphne M. Stam\textsuperscript{\afflink{15}},
Olivier Absil\textsuperscript{\afflink{16}},
\newauthor
Lisa Altinier\textsuperscript{\afflink{17}},
Pierre Baudoz\textsuperscript{\afflink{11}},
Ruslan Belikov\textsuperscript{\afflink{18}},
Alexis Bidot\textsuperscript{\afflink{19}},
Jayne L. Birkby\textsuperscript{\afflink{1}},
\newauthor
Markus J. Bonse\textsuperscript{\afflink{3}},
Bernhard Brandl\textsuperscript{\afflink{9}},
Alexis Carlotti\textsuperscript{\afflink{19}},
Elodie Choquet\textsuperscript{\afflink{17}},
Dirk van Dam\textsuperscript{\afflink{9}},
\newauthor
Niyati Desai\textsuperscript{\afflink{20}},
Kevin Fogarty\textsuperscript{\afflink{18}},
J. Fowler\textsuperscript{\afflink{21}},
Kyle van Gorkom\textsuperscript{\afflink{22}},
Yann Gutierrez\textsuperscript{\afflink{11},\afflink{23},\afflink{24}},
\newauthor
Olivier Guyon\textsuperscript{\afflink{22},\afflink{25},\afflink{26},\afflink{27}},
Sebastiaan Y. Haffert\textsuperscript{\afflink{22}},
Olivier Herscovici-Schiller\textsuperscript{\afflink{23}},
Adrien Hours\textsuperscript{\afflink{19}},
\newauthor
Roser Juanola-Parramon\textsuperscript{\afflink{28},\afflink{29}},
Evangelia Kleisioti\textsuperscript{\afflink{9},\afflink{30}},
Lorenzo König\textsuperscript{\afflink{16}},
Maaike van Kooten\textsuperscript{\afflink{31}},
\newauthor
Mariya Krasteva\textsuperscript{\afflink{32}},
Iva Laginja\textsuperscript{\afflink{11}},
Rico Landman\textsuperscript{\afflink{9}},
Lucie Leboulleux\textsuperscript{\afflink{19}},
David Mouillet\textsuperscript{\afflink{19}},
\newauthor
Mamadou N’Diaye\textsuperscript{\afflink{33}},
Emiel H. Por\textsuperscript{\afflink{34}},
Laurent Pueyo\textsuperscript{\afflink{34}},
Frans Snik\textsuperscript{\afflink{9}}
    \newauthor 
    \\%
    Affiliations are listed at the end of the paper
}
\date{Accepted 2023 July 07. Submitted 2023 May 26}

\begin{document} 

\maketitle

\begin{abstract}
NASA is engaged in planning for a Habitable Worlds Observatory (\HWO), a coronagraphic space mission to detect rocky planets in habitable zones and establish their habitability. 
Surface liquid water is central to the definition of planetary habitability.
Photometric and polarimetric phase curves of starlight reflected by an exoplanet can reveal ocean glint, rainbows and other phenomena caused by scattering by clouds or atmospheric gas.
Direct imaging missions are optimised for planets near quadrature, but \HWO' coronagraph may obscure the phase angles where such optical features are strongest. 
The range of accessible phase angles for a given exoplanet will depend on the planet's orbital inclination and/or the coronagraph's inner working angle (IWA). 
We use a recently-created catalog relevant to HabWorlds of 164 stars to estimate the number of exo-Earths that could be searched for ocean glint, rainbows, and polarization effects due to Rayleigh scattering. 
We find that the polarimetric Rayleigh scattering peak is accessible in most of the exo-Earth planetary systems.
The rainbow due to water clouds at phase angles of ${\sim}\qtyrange{20}{60}{\degree}$ would be accessible with \HWO\ for a planet with an Earth equivalent instellation in ${\sim}\num{46}$ systems, while the ocean glint signature at phase angles of ${\sim}\qtyrange{130}{170}{\degree}$ would be accessible in ${\sim}\num{16}$ systems, assuming an \IWA${=}\qty{62}{\mas}$ ($3\lambda/D$).
Improving the \IWA${=}\qty{41}{\mas}$ ($2\lambda/D$) increases accessibility to rainbows and glints by factors of approximately 2 and 3, respectively.
By observing these scattering features, \HWO could detect a surface ocean and water cycle, key indicators of habitability.
\looseness=-1
\end{abstract}

\begin{keywords}
    planets and satellites: terrestrial planets -- 
    instrumentation: high angular resolution -- 
    planets and satellites: atmospheres
\end{keywords}

%%%%%%%%%%%%%%%%%%%%%%%%%%%%%%%%%%%%%%%%%%%%%%%%%%%%%%%%%%%%%%%%%%%%%%%%%%%%%%%%%%%%%%%%%%%%%%%%%%%%%%%%%%%%%%%%%%%%%%%%%%%%%%
\section{Introduction}
\label{sec:intro}

%%%%%%%%%%%%%%%%%%%%%%%%%%%%%%%%%%%%%%%%%%%%%%%%%%%%%%%%%%%%%%%

Liquid surface water is closely related to planetary habitability because it is essential for life as we know it.
To first order, its presence can be predicted by the flux a planet receives from its star.
The habitable zone (HZ) is defined as the range of semi-major axes around a given main sequence star for which a rocky planet may have the right temperature to maintain liquid surface water \citep{kasting93}. 
However, there are many factors besides the incident stellar flux that influence a planet's ability to hold liquid surface water, such as the atmospheric pressure and geological activity. 
Detecting liquid surface water on an exoplanet would thus provide important context for the interpretation of possible biosignatures.
Variations of the flux and degree of polarisation as an exoplanet orbits its host star---phase variations---could reveal scattering phenomena from condensed water \citep[e.g., rainbows;][]{bailey2007} or specular reflection (i.e., glint) from water oceans \citep[e.g.][]{McCullough_2006,2008Icar..195..927W}.
However, depending on a planetary system's architecture, such features can be obscured by the inner working angle (\IWA) of a telescope's coronagraph, shown schematically in \cref{fig:annotated-orbit}.
In this paper, we quantify the planetary phase angles that would be accessible for direct observations of hypothetical terrestrial planets orbiting in the habitable zones of \HWO target stars.
\looseness=-1

%%%%%%%%%%%%%%%%%%%%%%%%%%%%%%%%%%%%%%%%%%%%%%%%%%%%%%%%%%%%%%%%%%%

\subsection{Angle conventions}

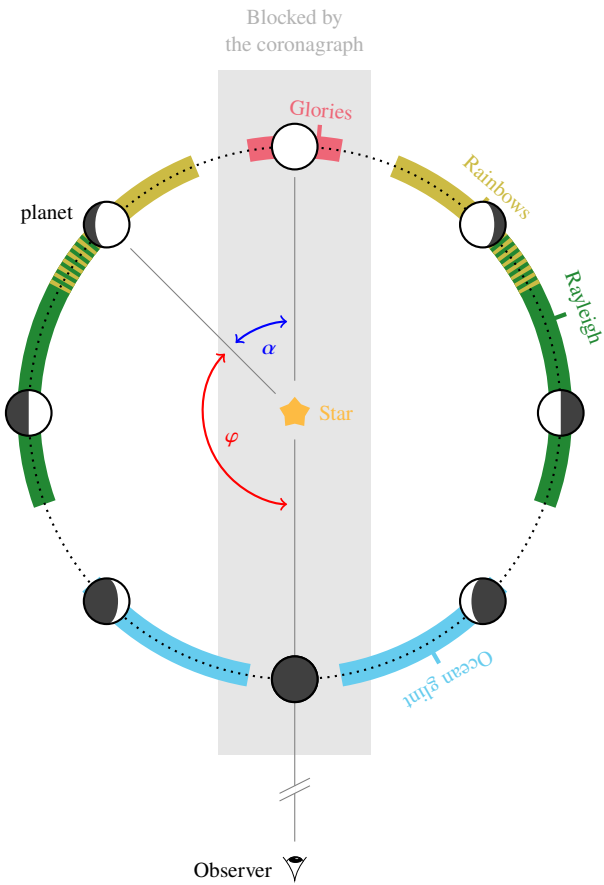
\begin{figure}
    \centering
    \begin{tikzpicture}

    % Define command for the "eye" symbol
    \newcommand*\lateraleye{%
       \scalebox{0.15}{
    \tikzset{every picture/.style={line width=0.75pt}}
    \begin{tikzpicture}[x=0.75pt,y=0.75pt,yscale=-1,xscale=1]
    \draw  [line width=1.5]  (300,100.33) .. controls (326,122) and (352,135) .. (378,139.33) .. controls (352,143.67) and (326,156.67) .. (300,178.33) ;
    \draw  [fill={rgb, 255:red, 0; green, 0; blue, 0 }  ,fill opacity=1 ] (308.94,116.33) .. controls (313.87,116.33) and (317.86,125.51) .. (317.85,136.83) .. controls (317.84,148.15) and (313.84,157.33) .. (308.91,157.33) .. controls (303.99,157.32) and (300,148.14) .. (300.01,136.82) .. controls (300.02,125.5) and (304.02,116.32) .. (308.94,116.33) -- cycle ;
    \draw  [draw opacity=0][line width=1.5]  (314.84,166.6) .. controls (311.87,164.64) and (309.14,162.18) .. (306.76,159.24) .. controls (295.12,144.82) and (296.6,124.33) .. (310.07,113.45) .. controls (311.48,112.32) and (312.96,111.33) .. (314.5,110.49) -- (331.14,139.55) -- cycle ; \draw  [line width=1.5]  (314.84,166.6) .. controls (311.87,164.64) and (309.14,162.18) .. (306.76,159.24) .. controls (295.12,144.82) and (296.6,124.33) .. (310.07,113.45) .. controls (311.48,112.32) and (312.96,111.33) .. (314.5,110.49) ;
    \draw  [fill={rgb, 255:red, 255; green, 255; blue, 255 }  ,fill opacity=1 ] (304.43,124.2) .. controls (306.09,124.25) and (307.32,128.01) .. (307.18,132.6) .. controls (307.05,137.19) and (305.59,140.88) .. (303.93,140.83) .. controls (302.27,140.78) and (301.03,137.02) .. (301.17,132.43) .. controls (301.31,127.83) and (302.76,124.15) .. (304.43,124.2) -- cycle ;
    \end{tikzpicture}
    }\,}

    % Command for planet dayside
    % Arguments: center, radius, color, rotation (-180,...,180)
    \newcommand\dayside[4]
    {%
      \ifnum #4 > 0
        \pgfmathsetmacro\lb{ #2} % left  arc, horizontal axis
        \pgfmathsetmacro\rb{ #2*(90-#4)/90} % right arc, horizontal axis
      \else
        \pgfmathsetmacro\lb{-#2*(90+#4)/90} % left  arc, horizontal axis
        \pgfmathsetmacro\rb{-#2} % right arc, horizontal axis
      \fi
      \draw[thick,#3,fill=white] #1 circle (#2);
      \fill[#3,opacity=0.75] ($#1+(0,#2)$) arc (90:270:\lb cm and #2 cm) arc (270:90:\rb cm and #2 cm);
    }

    % Define color scheme
    \definecolor{CBF1}{HTML}{EE6677} % blue
    \definecolor{CBF2}{HTML}{CCBB44} % orange
    \definecolor{CBF3}{HTML}{228833} % red
    \definecolor{CBF4}{HTML}{66CCEE} % purple

    % Background (to center stuff)
    \draw [fill=white, draw=none] (-3.3, 0) rectangle (3.3, 0);

    % Plot what's blocked by the coronagraph
    \draw [fill=black!10, draw=none] (-1, -4.5) rectangle (1, 4.5);
    \node [align=center, black!30, font=\small] at (0, 5) {Blocked by\\ the coronagraph};

    % Define radius
    \def\radius{3.5}

    % Command for arc with phase angle convention
    \def\phasearc[#1](#2)(#3:#4:#5){
        \begin{scope}[xscale=-1]
            \draw[#1, rotate=90] ($(#2)+({#5*cos(#3)},{#5*sin(#3)})$) arc (#3:#4:#5);
        \end{scope}
    }
    \newcommand{\phasenode}[3]{
        \node [#2, anchor=south, font=\small, rotate=-#1] at (90 - #1:\radius+0.3) {#3};
        \draw [#2, ultra thick] (90 - #1:\radius) -- (90 - #1:\radius+0.3);
    }

    % Color-code the orbit

    % Glories: (0, 5, 10)
    \phasearc [CBF1, line width=3mm](0, 0)(0:10:\radius);
    \phasearc [CBF1, line width=3mm](0, 0)(0:-10:\radius);
    \phasenode{5}{CBF1}{Glories};

    % Rainbow: (22, 42, 63)
    \phasearc [CBF2, line width=3mm](0, 0)(22:63:\radius);
    \phasearc [CBF2, line width=3mm](0, 0)(-22:-63:\radius);
    \phasenode{42}{CBF2}{Rainbows};

    % Rayleigh: (50, 70, 110)
    \phasearc [CBF3, line width=3mm](0, 0)(50:110:\radius);
    \phasearc [CBF3, line width=3mm](0, 0)(-50:-110:\radius);
    \phasenode{70}{CBF3}{Rayleigh};

    % Overlap between rainbows and Rayleigh
    \phasearc [CBF2, dash pattern=on 1.5pt off 1.5pt, dash phase=1.4pt, line width=3mm](0, 0)(50:63:\radius);
    \phasearc [CBF2, dash pattern=on 1.5pt off 1.5pt, dash phase=1.4pt, line width=3mm](0, 0)(-50:-63:\radius);

    % Glint: (130, 150, 170)
    \phasearc [CBF4, line width=3mm](0, 0)(130:170:\radius);
    \phasearc [CBF4, line width=3mm](0, 0)(-130:-170:\radius);
    \phasenode{150}{CBF4}{Ocean glint};

    % Plot the orbit
    \draw [dotted, black, thick] (0, 0) circle (\radius);

    % Star, planet position, and observer
    \node [star, fill=Dandelion, star points=5, inner sep=1mm] (star) (star) at (0, 0) {};
    \node [Dandelion, right=0.5mm of star] {Star};
    \node [] (planet) at (135:\radius+0.2) {};
    \node [] (seclipse) at (90:\radius) {};
    \node [black, left=1mm of planet.west] {planet};
    \node [circle, anchor=center, align=center, rotate=-90, inner sep=0mm] (observer) at (0, -6) {\lateraleye};
    \node [black, anchor=east, xshift=-2mm] at (observer) {Observer};

    % Plot some lines
    % \draw [shorten <= 1mm, shorten >= 3mm, densely dashed] (star) -- node [pos=0.45, font=\scriptsize, below=0mm, rotate=-45] {semi-major axis} (-45:\radius);
    \draw [shorten >= 5mm, shorten <= 2mm, gray] (star) -- (planet);
    \draw [shorten >= 3mm, shorten <= 2mm, gray] (star) -- (seclipse);
    \draw [shorten >= 1mm, shorten <= 2mm, gray] (star) -- (observer);
    \begin{scope}[yshift=-5cm]
        \draw [white, thick] (0, 0) -- (0, 0.1);
        \draw [gray] (-0.2, -0.1) -- (0.2, 0.1);
        \draw [gray] (-0.2, 0) -- (0.2, 0.2);
    \end{scope}

    % Plot planets (daysides)
    \foreach\i in {-180,-135,...,180} {
        \dayside{(\i+90:\radius)}{0.3}{black}{\i};
    }

    % Plot the scattering angle in red
    \draw [draw=red, thick, shorten >= 1mm, shorten <= 1mm, <->] ($(0, 0)+({1.2*cos(135)},{1.2*sin(135)})$) arc (135:270:1.2);
    \node [red] at (202.5:0.9) {$\varphi$};

    % Plot the phase angle in blue
    \draw [draw=blue, thick, shorten >= 1mm, shorten <= 1mm, <->] ($(0, 0)+({1.2*cos(90)},{1.2*sin(90)})$) arc (90:135:1.2);
    \node [blue] at (112.5:0.9) {$\alpha$};

\end{tikzpicture}
    \caption{
        Illustration of a planet in an edge-on, circular orbit showing how a coronagraph can obscure the exoplanet at small and large phase angles, and thus prevent observations of characteristic scattering features in an exoplanet's phase curve. 
        The scattering angle (the planet--star--observer angle), $\varphi$, is shown in red, and the supplementary phase angle, $\alpha$, is shown in blue. 
        See also \cref{fig:scattering-angle} and \cref{tab:phase_ranges}.
    }
    \label{fig:annotated-orbit}
\end{figure}

As shown in \cref{fig:annotated-orbit}, we define the \emph{scattering angle} $\varphi$ as the planet--star--observer angle (large angles for back-scattering, small angles for forward-scattering) and the supplementary \emph{phase angle} $\alpha$ (\qty{0}{\degree} at full phase and \qty{180}{\degree} at new phase). 
A planet on a perfectly face-on orbit is always seen at quadrature, $\varphi = \alpha = \qty{90}{\degree}$, while a planet on an inclined orbit will exhibit a wider range of viewing angles; only perfectly edge-on orbits pass through all possible scattering angles. 
In some plots, we also report \emph{degrees from quadrature}, so that both new and full phase are at \qty{90}{\degree} for edge-on orbits.

%%%%%%%%%%%%%%%%%%%%%%%%%%%%%%%%%%%%%%%%%%%%%%%%%%%%%%%%%%
% Figure 1
%%%%%%%%%%%%%%%%%%%%%%%%%%%%%%%%%%%%%%%%%%%%%%%%%%%%%%%%%%
%%%%%%%%%%%%%%%%%%%%%%%%%%%%%%%%%%%%%%%%%%%%%%%%%%%%%%%%%%

%%%%%%%%%%%%%%%%%%%%%%%%%%%%%%%%%%%%%%%%%%%%%%%%%%%%%%%%%%%%%%%
\subsection{Ocean glint}

Ocean glint is the specular (Fresnel) reflection of light off a smooth liquid surface. 
While glint is a familiar phenomenon on Earth, there is only one other body in the Solar System where it has been observed: the Cassini orbiter captured the glint of sunlight reflecting off the liquid hydrocarbon lakes near Titan's North Pole as it was observing Titan at crescent phases \citep{2010GeoRL..37.7104S}.
In the total flux, the glint signal is indeed expected to be most prominent at large planetary phase angles.
The maximum degree of polarisation of the glint occurs at the Brewster angle, which is determined by the refractive indices at the interface; between air and liquid water this occurs at a reflection angle of \qty{53}{\degree} locally on the surface \citep[at a planetary phase angle of about \qty{106}{\degree}; see, e.g.,][]{Zugger_2010,treesandstam2019}.
This angle therefore varies slightly with the wavelength and the composition of the ocean and the atmosphere.
The angular width of the glint in total and polarised flux generally broadens with increasing wave height \citep{2008Icar..195..927W, kopparla2018, treesandstam2019, trees2022}.
In general, the glint on a planet will be strongest at red wavelengths because there, Rayleigh scattering and attenuation in the atmosphere are smallest \citep{Robinson_2010,Zugger_2011}. 
This has been observed in Earthshine measurements \citep{Emde2017,sterzik2019, takahashi2021}. 
At large phase angles, the forward scattering of light by cloud particles could be confused with glint \citep{Robinson_2010}.  
This confusion can be resolved by measuring polarised phase curves at multiple wavelengths \citep{treesandstam2019}, as shown in \cref{fig:bottplot}, or by probing the rainbow, see below.
More recently, \citet{Ryan_Robinson_2022} have also proposed spectral PCA as a tool to separate glint from atmospheric effects on Earth-like worlds.

Glint provides evidence of specular reflecting surfaces, which, taken in the context of the temperature and types of condensates, may be interpreted as evidence for liquid water.
The presence of liquid water on the surface of a planet implies the presence of a water cycle, which provides key context to its habitability and the interpretation of atmospheric biosignatures.
Aside from detecting ocean glint, the presence of surface water on Earth-like exoplanets can be established through less direct methods near quadrature such as spectropolarimetry, to infer ocean reflection \citep{trees2022}, or through rotational mapping, to identify bright continents and dark oceans \citep[e.g.,][]{2009ApJ...700..915C,lustig2019}.

%%%%%%%%%%%%%%%%%%%%%%%%%%%%%%%%%%%%%%%%%%%%%%%%%%%%%%%%%%
% Figure 2
%%%%%%%%%%%%%%%%%%%%%%%%%%%%%%%%%%%%%%%%%%%%%%%%%%%%%%%%%%
\begin{figure*}
    \centering
    \includegraphics{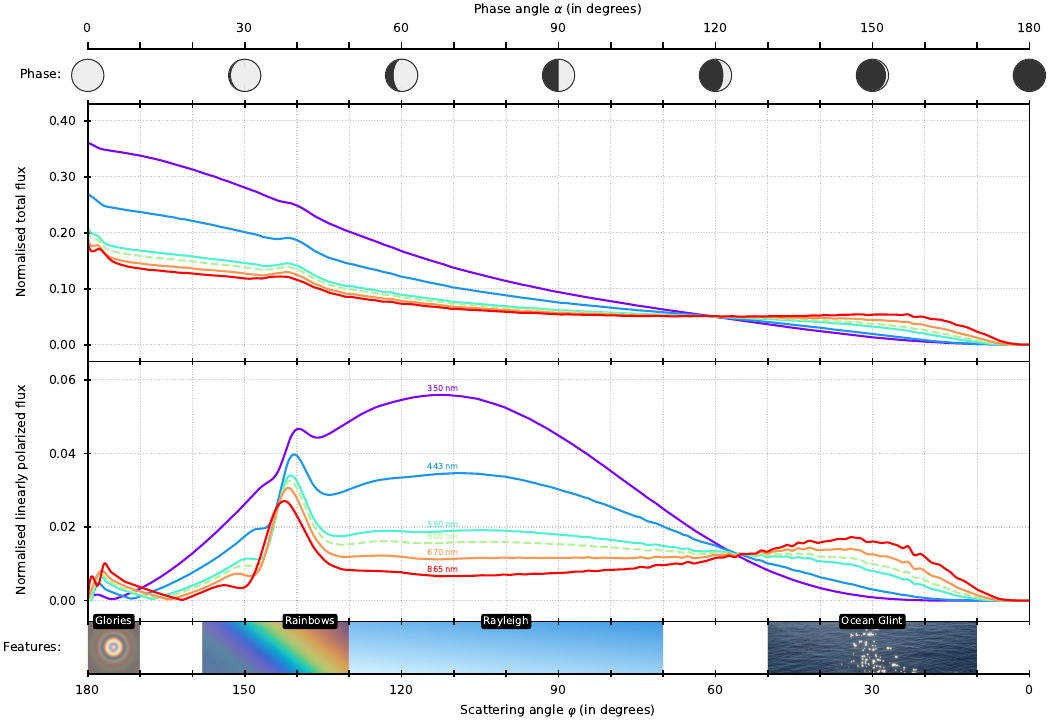}
    \script{create-figure-1-bott-plot.py}
    \caption{
        Phase curves of the normalised total and normalised linearly polarised flux of a cloudy planet with a wavy ocean surface and an Earth-like atmosphere in an edge-on orbit (adapted from \citealt{treesandstam2019}).  
        The wind speed over the ocean---which determines the wave height \citep{CoxMunk1954}---is \qty{7}{\meter\per\second}, and the clouds consist of spherical water droplets.
        The cloud coverage fraction is 0.5, and while the cloud pattern is patchy, the glint is always cloud-free. 
        The fluxes are normalised such that the total (i.e., unpolarised + polarised) flux equals the planet's geometric albedo at phase angle $\alpha = \qty{0}{\degree}$.
        In the bottom panel, we illustrate the scattering angle ranges of glories, rainbows, Rayleigh scattering, and ocean glint. 
        Note that the features in the actual curves partly overlap.
        The glory illustration is taken from \citet{Zakovryashin_2020}; all other illustrations are from \href{https://unsplash.com}{unsplash.com} [\href{https://unsplash.com/photos/YWL4dqqiRdk}{1}, \href{https://unsplash.com/photos/BgFOMcg4HNs}{2}, \href{https://unsplash.com/photos/9YdePM6BrO4}{3}].
        This figure is best viewed in colour.
    }
    \label{fig:bottplot}
\end{figure*}
%%%%%%%%%%%%%%%%%%%%%%%%%%%%%%%%%%%%%%%%%%%%%%%%%%%%%%%%%%%%%%%

%%%%%%%%%%%%%%%%%%%%%%%%%%%%%%%%%%%%%%%%%%%%%%%%%%%%%%%%%%
\subsection{Rainbows, cloudbows, and glories}

Rainbows, cloudbows, glories, and other scattering phenomena can dramatically influence the shape of a planet's photometric and polarimetric phase curves and provide a wealth of information about exoplanets \citep{karalidi2012rainbow, stam2008,bailey2007,2014A&A...566L...1G}.%
\footnote{
    Rainbows occur when light is scattered by rain droplets that are large compared to the wavelength of the light. 
    Cloudbows refer to features occurring at the same phase angles from smaller cloud particles, with the spread in colours invisible to the human eye making them appear white \citep[for further discussion, see][]{bailey2007}.  
}
In the Solar System, the polarised back-scatter from Titan indicated that the moon's reflection was due to an atmospheric haze rather than a surface \citep{zellner1973polarization}, and Venus's polarised phase curve at three wavelengths showed that its clouds were a $\sim\qty{75}{\percent}$ H\textsubscript{2}SO\textsubscript{4}--water mixture and constrained the droplet size distribution \citep{hansenhovenier1974}.
The glory feature in Venus's flux phase curve was later used to rule out various species as the unknown UV absorber \citep{petrova2018glory}.
Glories and rainbows are technically visible both in total and polarised flux phase curves in the presence of spherical cloud and/or haze particles, but their strongly polarised nature coupled with the suppression of other polarised features at gibbous and full phases make them more easily characterized with polarimetry than with photometry \citep{karalidi2011, stam2008, treesandstam2019,trees2022}.
\looseness=-1

The phase angles and wavelength dependence of these features are sensitive to the chemical composition and droplet size distribution of condensates in the atmosphere. 
As with ocean glint, the detection of liquid condensates on an exoplanet can provide key context to their habitability, potential water cycle, and interpretation of atmospheric biosignatures.
In this paper, we define our optimal angle for the detection of a rainbow as that of the peak of the water droplet rainbow at visible wavelengths, which is around a phase angle $\alpha$ of \qty{40}{\degree} (see \cref{fig:bottplot}).
Pushing observations to angles closer to the star (smaller and larger phase angles) would allow the determination of the full shape of the rainbow, and enable the identification of rainbows due to the scattering of light by other condensates.
The glory can allow further characterization of condensates as its angle and shape also depend on the refractive index, but this glory occurs within a phase angle of \qty{20}{\degree}, when the on-sky separation is small, as can also be seen in \cref{fig:bottplot}. 

%%%%%%%%%%%%%%%%%%%%%%%%%%%%%%%%%%%%%%%%%%%%%%%%%%%%%%%%%%%%%%%
\subsection{Habitable Worlds Observatory}

The National Academy of Sciences Astronomy \& Astrophysics 2020 Decadal Survey \citep[][hereafter \Decadal]{decadal} recommended a telescope with the capability to detect signatures of habitability on approximately 25 HZ planets in a new \enquote{Great Observatories} program.
This requires an instrument with a coronagraph capable of high-contrast imaging and low-resolution spectroscopy at optical to near-infrared wavelengths. 
The precursor technology recommended by the survey lists \enquote{direct imaging to probe polarised ocean glint on terrestrial planets} as a priority capability \citep[Table E.1 in][]{decadal} for ground and space-based observatories.
Following the release of this report, NASA recently announced the start of the development of the Habitable Worlds Observatory.
The performance and characteristics (e.g., on- or off-axis, diameter, type of segmentation, polarimetry or photometry, coronagraph, \IWA) of \HWO will significantly impact the expected exo-Earth yield \citep{Stark2019_exoplanetyield} and are still to be determined.
They will likely be heavily informed by the LUVOIR \citep{LUVOIR2019} and HabEx \citep{HabEx_2020} preparatory studies.
For generality, in this work, we report angular separations in milliarcseconds; however, for ease of comparison, \cref{tab:IWA_OWA} lists the angular separations corresponding to multiples of $\lambda / D$ for a $\qty{6}{\meter}$ telescope, as suggested in \Decadal, at representative wavelengths $\lambda$ of $\qty{600}{\nano\meter}$ and $\qty{1}{\micro\meter}$.

%%%%%%%%%%%%%%%%%%%%%%%%%%%%%%%%%%%%%%%%%%%%%%%%%%%%%%%%%%
% Table 1
%%%%%%%%%%%%%%%%%%%%%%%%%%%%%%%%%%%%%%%%%%%%%%%%%%%%%%%%%%
\begin{table}
    \centering
    \caption{
        Inner and Outer Working Angles (\IWA and \OWA{}, respectively) in milliarcseconds (\unit{\mas}) as functions of $\lambda / D$, at wavelengths $\lambda$ of \qty{600}{\nano\meter} and \qty{1}{\micro\meter}, assuming a telescope diameter $D$ of \qty{6}{\meter}.
    }
    \label{tab:IWA_OWA}
    \script{create-table-1-iwa-owa.py}
    
% ATTENTION:
% THIS FILE IS GENERATED AUTOMATICALLY BY `create-table-1-iwa-owa.py`
% PLEASE DO NOT EDIT MANUALLY

% LAST UPDATE: 2023-05-25T23:56:01.098903 
% GENERATED ON: Timothys-MBP 

\begin{tabular}{rrrr}
\toprule
     &   $\lambda / D$ &   mas (at \qty{600}{\nano\meter}) &   mas (at \qty{1}{\micro\meter}) \\
\midrule
 IWA &               1 &                              20.6 &                             34.4 \\
 IWA &               2 &                              41.3 &                             68.8 \\
 IWA &               3 &                              61.9 &                            103.1 \\
 IWA &               4 &                              82.5 &                            137.5 \\
\midrule
 OWA &              32 &                             660.0 &                           1100.1 \\
 OWA &              64 &                            1320.1 &                           2200.2 \\
\bottomrule
\end{tabular}

\end{table}
%%%%%%%%%%%%%%%%%%%%%%%%%%%%%%%%%%%%%%%%%%%%%%%%%%%%%%%%%% 
 
%%%%%%%%%%%%%%%%%%%%%%%%%%%%%%%%%%%%%%%%%%%%%%%%%%%%%%%%%%%%%%%%%%%%%%%%%%%%%%%%%%%%%%%%%%%%%%%%%%%%%%%%%%%%%%%%%%%%%%%%%%%%%%
\section{Methods}

%%%%%%%%%%%%%%%%%%%%%%%%%%%%%%%%%%%%%%%%%%%%%%%%%%%%%%%%%%%%%%%
\subsection{Target list}
\label{subsec:2.1}

A preliminary target list of nearby stars (maximum distance \qty{25}{\parsec}) around which \HWO may search for habitable rocky planets (assuming a \qty{6}{\meter} telescope), has been compiled: the \emph{NASA Exoplanet Exploration Program's Mission Star List for the Habitable Worlds Observatory}.%
\footnote{Available online: \url{https://exoplanets.nasa.gov/internal_resources/2645_NASA_ExEP_Target_List_HWO_Documentation_2023.pdf}}
This list consists of $\num{164}$ stars, the majority of which are Sun-like dwarfs: 66~F~dwarfs, 55~G~dwarfs, 40~K~dwarfs, and 3~M~dwarfs.
The target list is constructed assuming a maximum planet magnitude in the Cousins $R_c$ band of $31$ mag and a lower limit of \num{2.5e-11} for the planet-star flux ratio. 
The adopted HZ limits are a semi-major axis of \qtyrange{0.95}{1.67}{\au} for a solar twin and planet sizes between \qtyrange{0.8}{1.4}{} Earth radii. 
For non-solar stars, the semi-major axis scales as the square root of the bolometric luminosity normalized to the Sun. 
This range of orbital separation corresponds to the \enquote{conservative habitable zone} \citep{kasting93, kopparapu13}. 
In the remainder of this work, we consider hypothetical Earth-sized planets orbiting at the right distance from their parent star to receive the same flux as the Earth receives from the Sun.
By design, all such planets should be observable by \HWO at quadrature phase. 
\looseness=-1

%%%%%%%%%%%%%%%%%%%%%%%%%%%%%%%%%%%%%%%%%%%%%%%%%%%%%%%%%%%%%%%

\subsection{Accessible phase angles}
\label{subsec:2.2}

We first define the phase angle ranges over which the various phenomena---glint, Rayleigh scattering peak, rainbow, and glory---manifest themselves in \cref{tab:phase_ranges}, also shown in \cref{fig:bottplot}. 
These ranges are all approximate, as the exact angles where these phenomena dominate depend on the details of the atmospheric and surface properties, as well as on the wavelengths being observed.

%%%%%%%%%%%%%%%%%%%%%%%%%%%%%%%%%%%%%%%%%%%%%%%%%%%%%%%%%%
% Table 2
%%%%%%%%%%%%%%%%%%%%%%%%%%%%%%%%%%%%%%%%%%%%%%%%%%%%%%%%%%
\begin{table}
    \centering
    \caption{
        Approximate ranges of the phase angle $\alpha$ for the optical phenomena considered in this paper (cf. \cref{fig:bottplot}). 
        The scattering feature is present between $\alpha_\mathrm{min}$ and $\alpha_\mathrm{max}$; $\alpha_\mathrm{peak}$ is the phase angle where the scattered flux peaks.
        \looseness=-1
    }
    \label{tab:phase_ranges}
    \begin{tabular}{ l c c c } 
        \toprule
        & $\alpha_\mathrm{min}$ & $\alpha_\mathrm{peak}$ & $\alpha_\mathrm{max}$ \\
        \midrule
        Glory       & \qty{0}{\degree}     & \qty{5}{\degree}     & \qty{10}{\degree} \\
        Rainbow    & \qty{22}{\degree}    & \qty{42}{\degree}    & \qty{63}{\degree} \\
        Rayleigh peak    & \qty{50}{\degree}    & \qty{70}{\degree}    & \qty{110}{\degree} \\
        Ocean Glint & \qty{130}{\degree}   & \qty{150}{\degree}   & \qty{170}{\degree} \\
        \bottomrule
    \end{tabular}
\end{table}
%%%%%%%%%%%%%%%%%%%%%%%%%%%%%%%%%%%%%%%%%%%%%%%%%%%%%%%%%%

A coronagraphic mask can obscure part of a planet's orbit. 
As illustrated in \cref{fig:annotated-orbit}, obscuration is a problem at small and large phase angles, when the planet is at small angular separation from its star.
The range of accessible phase angles will depend on the \IWA of the coronagraph, as well as on the orbital inclination and eccentricity.

%%%%%%%%%%%%%%%%%%%%%%%%%%%%%%%%%%%%%%%%%%%%%%%%%%%%%%%%%%%%%%%
\subsubsection{Circular, edge-on orbits}
 
\begin{figure}
    \centering
    \begin{tikzpicture}

    % Define command for the "eye" symbol
    \newcommand*\lateraleye{%
       \scalebox{0.15}{
    \tikzset{every picture/.style={line width=0.75pt}} 
    \begin{tikzpicture}[x=0.75pt,y=0.75pt,yscale=-1,xscale=1]
    \draw  [line width=1.5]  (300,100.33) .. controls (326,122) and (352,135) .. (378,139.33) .. controls (352,143.67) and (326,156.67) .. (300,178.33) ;
    \draw  [fill={rgb, 255:red, 0; green, 0; blue, 0 }  ,fill opacity=1 ] (308.94,116.33) .. controls (313.87,116.33) and (317.86,125.51) .. (317.85,136.83) .. controls (317.84,148.15) and (313.84,157.33) .. (308.91,157.33) .. controls (303.99,157.32) and (300,148.14) .. (300.01,136.82) .. controls (300.02,125.5) and (304.02,116.32) .. (308.94,116.33) -- cycle ;
    \draw  [draw opacity=0][line width=1.5]  (314.84,166.6) .. controls (311.87,164.64) and (309.14,162.18) .. (306.76,159.24) .. controls (295.12,144.82) and (296.6,124.33) .. (310.07,113.45) .. controls (311.48,112.32) and (312.96,111.33) .. (314.5,110.49) -- (331.14,139.55) -- cycle ; \draw  [line width=1.5]  (314.84,166.6) .. controls (311.87,164.64) and (309.14,162.18) .. (306.76,159.24) .. controls (295.12,144.82) and (296.6,124.33) .. (310.07,113.45) .. controls (311.48,112.32) and (312.96,111.33) .. (314.5,110.49) ;
    \draw  [fill={rgb, 255:red, 255; green, 255; blue, 255 }  ,fill opacity=1 ] (304.43,124.2) .. controls (306.09,124.25) and (307.32,128.01) .. (307.18,132.6) .. controls (307.05,137.19) and (305.59,140.88) .. (303.93,140.83) .. controls (302.27,140.78) and (301.03,137.02) .. (301.17,132.43) .. controls (301.31,127.83) and (302.76,124.15) .. (304.43,124.2) -- cycle ;
    \end{tikzpicture}
    }\,}

    % Star
    \node [star, fill=Dandelion, star points=5, inner sep=1mm] (star) at (0, 0) {};
    \node [below=0mm of star, font=\scriptsize] {Star};

    % Planet
    \node [circle, fill=ForestGreen, inner sep=0.5mm] (planet) at (-1.5, -1.5) {};
    \node [below=0mm of planet, font=\scriptsize] {Planet};

    % Observer
    \node [inner sep=0mm] (observer) at (-7, 0) {\reflectbox{\lateraleye}};
    \node [below=0mm of observer, font=\scriptsize] {Observer};

    \draw [shorten >= 1mm, shorten <= 1mm, <->] (planet) -- node [pos=0.5, left=0mm, font=\scriptsize] {$r_\mathrm{proj}$} ++(0, 1.5);

    % phi
    \draw[draw=red, <->, shorten >= 0.5mm, shorten <= 0.5mm] (-0.75, 0) arc[start angle=180, end angle=225, radius=0.75cm] node [pos=0.5, red, font=\scriptsize, xshift=-2mm, yshift=-1mm] {$\varphi$};

    \draw[draw=Cerulean, <->, shorten >= 0.5mm, shorten <= 0.5mm] (-5.5, 0) arc[start angle=0, end angle=-23, radius=1cm] node [pos=0.5, Cerulean, font=\scriptsize, xshift=3mm, yshift=-0.25mm] {$\delta$};

    \draw [shorten >= 1mm, shorten <= 1mm, densely dotted] (star) --node [pos=0.6, right=0.1, font=\scriptsize] {$a$} (planet);
    \draw [shorten >= 1mm, shorten <= 1mm, densely dotted] (star) -- node [pos=0.5, above=0, font=\scriptsize] {$d_*$} (observer);
    \draw [shorten >= 1mm, shorten <= 1mm, densely dotted] (planet) -- (observer);

\end{tikzpicture}
    \caption{
        The angular separation of a planet from its star as seen by the observer, $\delta$, is related to the scattering angle $\varphi$. 
        The star is at a distance $d_*$ from the observer, the planet orbits a distance $a$ from the star, and  the projected distance between the star and planet is  $r_\mathrm{proj}$. 
        Not to scale. 
    }
    \label{fig:scattering-angle}
\end{figure}
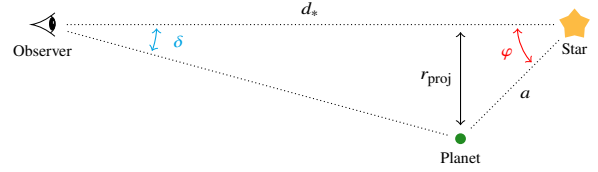

The on-sky projected planet distance, $r_\mathrm{proj}$ (in \unit{\au}) is shown in \cref{fig:scattering-angle} for a non-eccentric case with a semi-major axis $a$ (in \unit{\au}), at a distance $d_*$ (in \unit{\parsec}) from the observer.
The scattering angle, $\varphi$, is given by
\begin{equation}
    \label{eq:sinphi}
    \sin(\varphi) = - \sin(\alpha) = \frac{r_\mathrm{proj}}{a} \,,
\end{equation}
assuming small angles and $r_\mathrm{proj} = \delta d_*$, where $\delta$ is the on-sky angular separation.
We access the minimum/maximum phase angle when $r_\mathrm{proj}$ reaches the inner working angle separation, thus when $\delta = \mathrm{IWA}$. 
We can rewrite \cref{eq:sinphi} as:
\begin{equation}
    \label{eq:scattering_angle}
    \cos\left(\dfrac{\Delta \alpha}{2}\right) = \frac{\mathrm{IWA} 
    \cdot d_*}{a} \,.
\end{equation}
where we define $\Delta \alpha$ as the difference between the maximum and 
minimum accessible phase angles. For non-eccentric orbits the maximum and minimum phase angles will be equal sides of quadrature and so $\Delta \alpha / 2$ represents the maximum angle away from quadrature that can be accessed.

%%%%%%%%%%%%%%%%%%%%%%%%%%%%%%%%%%%%%%%%%%%%%%%%%%%%%%%%%%
% Figure 3
%%%%%%%%%%%%%%%%%%%%%%%%%%%%%%%%%%%%%%%%%%%%%%%%%%%%%%%%%%
%%%%%%%%%%%%%%%%%%%%%%%%%%%%%%%%%%%%%%%%%%%%%%%%%%%%%%%%%%

\begin{figure*}
    \centering
    \includegraphics{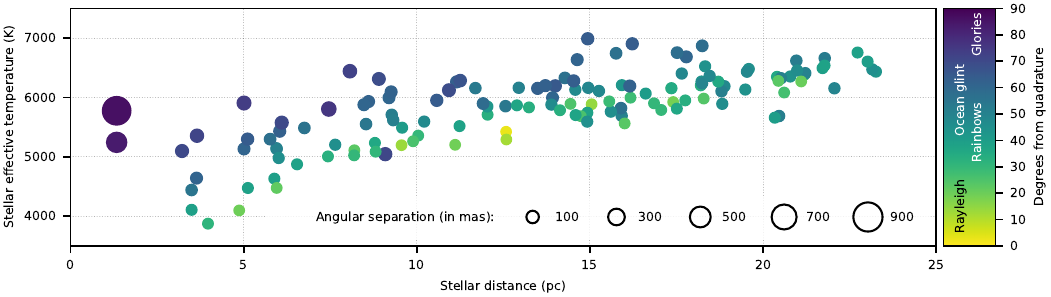}
    \script{create-figure-4-scatterplot.py}
    \caption{
        Scatter plot for the \HWO target stars, showing stellar effective temperatures and distances.
        The colours show $\Delta \alpha$ for hypothetical planets on circular, edge-on orbits at a semi-major axis $a$ corresponding to an Earth-like instellation and an IWA of \qty{61.9}{\mas}.
        The colour bar indicates the optical phenomena that can in principle be detected: all the phenomena from the bottom of the bar up to the colour of a given circle would be detectable for that planet.
        For example, dark blue circles indicate planets for which most phenomena would be accessible: planets that can be observed at the rainbow angle can also be observed at angles exhibiting the polarization peak due to Rayleigh scattering.
    }
    \label{fig:scatterplot}
\end{figure*}

\Cref{fig:scatterplot} shows {\bf $\Delta \alpha / 2$} for the systems in the \HWO target list, assuming an IWA of \qty{61.9}{\mas}.
As can be seen in \cref{fig:scatterplot}, for exoplanets in edge-on circular orbits, it will always be possible to observe the peak polarisation due to Rayleigh scattering, which occurs around quadrature ($\alpha=\qty{90}{\degree}$), while the rainbow angle ($\alpha\sim\qty{40}{\degree}$) will only be accessible for a subset of exoplanets.

%%%%%%%%%%%%%%%%%%%%%%%%%%%%%%%%%%%%%%%%%%%%%%%%%%%%%%%%%%
% Figure 4
%%%%%%%%%%%%%%%%%%%%%%%%%%%%%%%%%%%%%%%%%%%%%%%%%%%%%%%%%%
%%%%%%%%%%%%%%%%%%%%%%%%%%%%%%%%%%%%%%%%%%%%%%%%%%%%%%%%%%

%%%%%%%%%%%%%%%%%%%%%%%%%%%%%%%%%%%%%%%%%%%%%%%%%%%%%%%%%%%%%%%
\subsubsection{Circular, inclined orbits}

Most planets are not in edge-on orbits as seen from Earth.
For an orbit with an inclination angle $i$ (where $i=\qty{0}{\degree}$ is face-on), we can identify the following two regimes:
\begin{enumerate}
    \item The coronagraph does not obscure any part of the sky-projected orbit: 
    the maximum phase angle coverage depends solely on $i$. 
        This holds for orbits that are relatively face-on (small $i$). 
    \item The coronagraph does obscure parts of the sky-projected orbit: the phase angle coverage depends only on the \IWA and the orbital semi-major axis $a$, following \cref{eq:scattering_angle}. 
\end{enumerate}

The range of accessible phase angles is then 
\begin{equation}
\label{eq:Delta_phi_max}
    \Delta \alpha = 
    \begin{cases}
        2 i & \textrm{for} \cos(i) > \frac{\mathrm{IWA} \cdot d_* }{a} \\ 
        2 \cos^{-1}\left(\dfrac{\mathrm{IWA} \cdot d_* }{a}\right) & \textrm{for} \cos(i) < \frac{\mathrm{IWA} \cdot d_* }{a}
    \end{cases} \,
\end{equation}

%%%%%%%%%%%%%%%%%%%%%%%%%%%%%%%%%%%%%%%%%%%%%%%%%%%%%%%%%%%%%%%
\subsection{Monte Carlo simulations of accessible orbits}
\label{subsec:2.3}

\subsubsection{Circular, randomly inclined orbits}
\label{sec:circular}

\begin{figure*}%[t]
    \centering
    \includegraphics{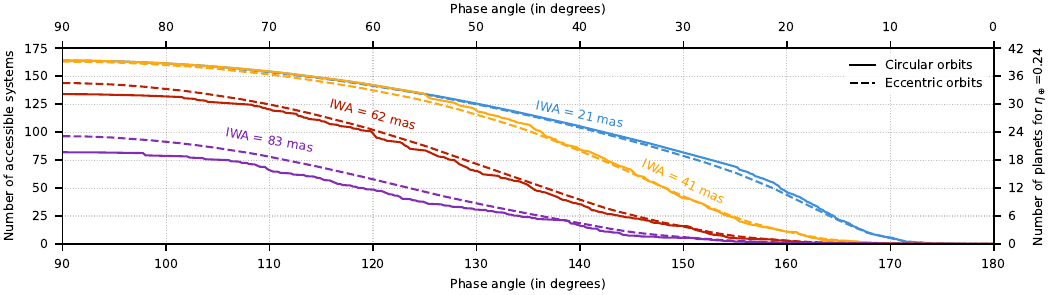}
    \script{create-figure-5-n-over-phase.py}
    \caption{
        Expected number (computed as the average over \num{1000} simulations) of the most extreme phase angles accessible for different \IWA and for randomly inclined, circular orbits (solid lines) and randomly inclined, eccentric orbits (dashed lines).
        The top and bottom $x$-axes indicate the minimum and maximum accessible phase angles, respectively.
        These angles are symmetric about quadrature (\qty{90}{\degree}).
        The $y$-axis on the left indicates the number of planetary systems divided by the number of Monte Carlo samples, and is thus normalised to the number of systems on the target list.
        The $y$-axis on the right indicates the number of HZ Earth-like planets that could be imaged at these phase angles, assuming an occurrence rate of $\eta_\oplus = \qty{24}{\percent}$ (as in \Decadal).
    }
    \label{fig:betaallofit}
\end{figure*}

For each of the 164 stars on the \HWO target list, we generate \num{1000} random orbits of hypothetical habitable planets with semi-major axes that yield an Earth-equivalent-incident stellar flux. 
The orbital inclination angles $i$ are randomly drawn from a distribution that is uniform in $\cos i$. 
We then use \cref{eq:Delta_phi_max} to determine the range of accessible phase angles for each hypothetical planet.
The solid lines in \cref{fig:betaallofit} show the cumulative distribution of the maximum and minimum accessible phase angles normalised to the total number of systems in the \HWO target list for different values of the \IWA.

%%%%%%%%%%%%%%%%%%%%%%%%%%%%%%%%%%%%%%%%%%%%%%%%%%%%%%%%%%
% Figure 5
%%%%%%%%%%%%%%%%%%%%%%%%%%%%%%%%%%%%%%%%%%%%%%%%%%%%%%%%%%
%%%%%%%%%%%%%%%%%%%%%%%%%%%%%%%%%%%%%%%%%%%%%%%%%%%%%%%%%%

%%%%%%%%%%%%%%%%%%%%%%%%%%%%%%%%%%%%%%%%%%%%%%%%%%%%%%%%%%
% Figure 6: Yarn ball
%%%%%%%%%%%%%%%%%%%%%%%%%%%%%%%%%%%%%%%%%%%%%%%%%%%%%%%%%%
\begin{figure*}
    \centering
    \includegraphics{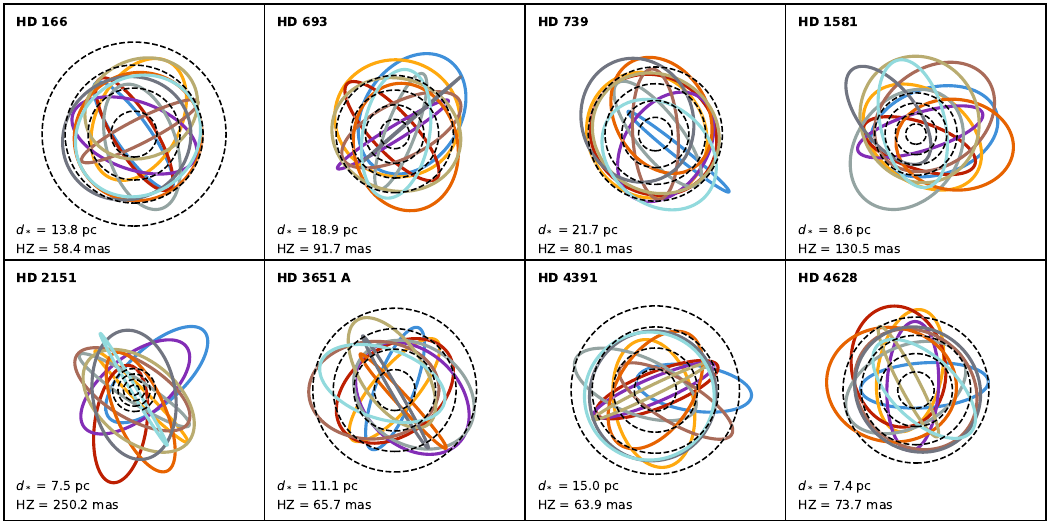}  
    \script{create-figure-6-yarnball.py}
    \caption{
        Examples of the eccentric orbits generated for the stellar sample. 
        Each plot shows 10 examples of orbits randomly drawn for each star.
        The orbits are scaled to ensure an Earth-like incident flux.
        The concentric, dashed circles indicate inner working angles of \num{20.6}, \num{41.3}, \num{61.9}, and \qty{82.5}{\mas}.
        Each system has been scaled such that its HZ has the same radius, hence the variation in the values of the \IWA{}s compared to the orbits.
        The plot clearly shows that the \IWA can significantly affect the range of accessible phases angles for each orbit.
    }
    \label{fig:ball-o-yarn}
\end{figure*}
%%%%%%%%%%%%%%%%%%%%%%%%%%%%%%%%%%%%%%%%%%%%%%%%%%%%%%%%%%

%%%%%%%%%%%%%%%%%%%%%%%%%%%%%%%%%%%%%%%%%%%%%%%%%%%%%%%%%%%%%%%
\subsubsection{Eccentric, randomly inclined orbits}
\label{sec:eccentric}

We repeat the Monte Carlo simulations from \cref{sec:circular}, assuming the orbital eccentricity follows a beta-distribution with shape parameters $a=0.867$ and $b=3.03$ \citep{2013MNRAS.434L..51K}. 
The semi-major axis is kept the same for the circular and eccentric orbits---we do not renormalise the semi-major axis to keep the time-averaged flux the same for all systems.
We randomly draw \num{1000} orbits for each star and determine the sky-projected orbits following, for example, \citet{2010exop.book...15M}. 
Orbital eccentricities are drawn from the beta distribution, and inclination angles are uniform in $\cos i$. 
\Cref{fig:ball-o-yarn} shows a sample of the generated orbits.

We then calculate the minimum and maximum accessible phase angles for \IWA{}s of $\qtylist{21; 41; 62; 83}{\mas}$.
The dashed lines in \cref{fig:betaallofit} show the normalised cumulative distribution of the accessible phase angles for each value of the \IWA for eccentric orbits. 
We also calculate which target planets would be accessible at the phase angles where ocean glint, the rainbow, or the Rayleigh scattering peak occur.
\looseness=-1

%%%%%%%%%%%%%%%%%%%%%%%%%%%%%%%%%%%%%%%%%%%%%%%%%%%%%%%%%%%%%%%
\subsection{Simulating contrast curves}
\label{subsec:2.4}

A second parameter that is crucial for detecting angular features like rainbows and ocean glint on Earth-like exoplanets is the contrast, that is, the planet-to-star flux ratio. 
While contrast is not the main focus of this paper, we note that the total and polarised flux contrasts depend on the phase angle, due to a combination of the changing illumination fraction of the observable planetary disk and non-isotropic reflection by the planet.
For the contrast calculations, we start with the \HWO target list (see \cref{subsec:2.1}), which, for every star, includes an estimated contrast for an Earth-twin in the habitable zone.
The contrast is given by \citep[e.g.,][]{2023A&A...671A.165M}
\begin{equation}
    C(\alpha) 
    = F_p(\alpha) / F_* 
    = A_\mathrm{G} \cdot \phi(\alpha) \cdot \left( R_p / a(\alpha) \right)^2 \,,
    \label{eq:contrast}
\end{equation}
where $A_\mathrm{G}$ is the planet's geometric albedo, $R_p$ its radius, $\phi(\alpha)$ is the phase function at phase angle $\alpha$ (normalized to \num{1} at full phase, thus $\phi(0^\circ)= 1.0$, and $a$ is the distance between the planet and its host star.

The contrast estimations in the \HWO target list were 
calculated assuming circular planetary orbits, and Lambertian reflecting planets \citep{Russell1916} with a geometric albedo $A_{\rm G}$ equal to 0.2.
This albedo value was adopted from the VPL models in \citet{Robinson2011}, and the assumption of a Lambert phase function is justifiable given the apparent albedo values derived from Earthshine observations by \citet{Qiu2003} and \citet{Palle2003}.
For a more accurate representation of the contrast as a function of the separation, we replace the Lambertian phase function with the models shown in \cref{fig:bottplot} from \citet{treesandstam2019} which assume an Earth-like planet with an ocean surface, a wind-speed of $\qty{7}{\meter\per\second}$, and patchy liquid water clouds observed at $\lambda = \qty{670}{\nano\meter}$.\looseness=-1 

\begin{figure*}
    \centering
    \includegraphics{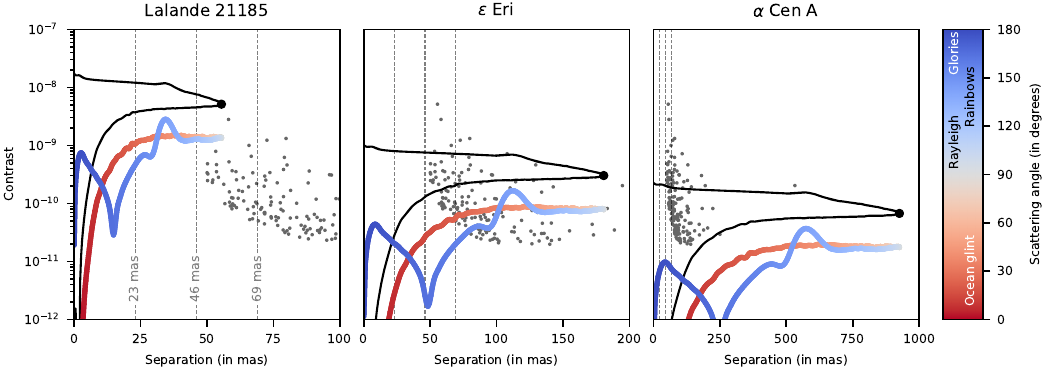}
    \script{create-figure-7-contrast-over-separation.py}
    \caption{
        Planet-to-star contrast ratio and orbital separation for an Earth-like planet with an ocean and patchy clouds along its circular orbit  around Lalande~21185 (M2V), $\epsilon$~Eri (K2V), and $\alpha$~Cen~A (G2V) when assuming an inclination of $i = \qty{90}{\degree}$.
        The solid black lines indicate the contrast in unpolarised light at $\lambda = \qty{670}{\nano\meter}$, with the contrast at quadrature marked by a black dot.
        The contrast for linearly polarised light at $\lambda = \qty{670}{\nano\meter}$ is indicated by the coloured lines, with the colour encoding the scattering angle.
        The small grey dots show the contrasts at quadrature of the other targets in the star list, and the dashed lines indicate \IWA's of $\qtylist{23; 46; 69}{\mas}$ (corresponding to 1, 2, 3 $\lambda /D $ at $\lambda = \qty{670}{\nano\meter}$).
    }
    \label{fig:contrasts}
\end{figure*}

We assume that the contrasts in the \HWO target list represent the reflected flux at quadrature ($\alpha = \qty{90}{\degree}$), where a Lambertian phase function equals $1/\pi$. 
Using \cref{eq:contrast}, we correct the contrasts in the \HWO target list by dividing them with $0.2/\pi$ (=\,\qty{6.37}{\percent}) and multiplying them by the  reflected light fractions of the updated models of \citet{treesandstam2019}.
We then map the orbital phase for circular orbits to the phase angle and on-sky separation, assuming an inclination $i$ of \qty{90}{\degree}. 
Together, this gives the reflected light contrast as a function of separation for a full orbit.
We multiply this flux with the degree of polarisation to obtain the polarised contrast.
\Cref{fig:contrasts} shows the resulting contrast curves for three of the closest target stars: the M~dwarf Lalande~21185, the K~dwarf $\epsilon$~Eri, and the G~dwarf $\alpha$~Cen~A.
These calculations are by no means complete or exact, given the many variables that go into the modelling of reflected total and polarised fluxes of Earth-like exoplanets \citep{treesandstam2019,trees2022}.

%%%%%%%%%%%%%%%%%%%%%%%%%%%%%%%%%%%%%%%%%%%%%%%%%%%%%%%%%%
% Figure 7
%%%%%%%%%%%%%%%%%%%%%%%%%%%%%%%%%%%%%%%%%%%%%%%%%%%%%%%%%%
%%%%%%%%%%%%%%%%%%%%%%%%%%%%%%%%%%%%%%%%%%%%%%%%%%%%%%%%%%

%%%%%%%%%%%%%%%%%%%%%%%%%%%%%%%%%%%%%%%%%%%%%%%%%%%%%%%%%%%%%%%%%%%%%%%%%%%%%%%%%%%%%%%%%%%%%%%%%%%%%%%%%%%%%%%%%%%%%%%%%%%%%%

\vspace{-1mm}
\section{Results}
\label{sec:3}

%%%%%%%%%%%%%%%%%%%%%%%%%%%%%%%%%%%%%%%%%%%%%%%%%%%%%%%%%%%%%%%

\subsection{The effect of eccentricity}
\label{sec:result_eccentricity}

As shown in \cref{fig:betaallofit}, an empirical eccentricity distribution increases the number of systems that can be observed at each phase angle (notably for the red and purple lines). 
This is because eccentric planets reach orbital separations greater than their semi-major axis, and hence are more likely to \enquote{peek out} from behind the coronagraph \IWA at extreme orbital phases (albeit with worse contrast due to the greater star-planet separation). 
The eccentricity does not significantly affect the number of systems for \IWA smaller than that used in the selection of the \HWO target list (yellow and blue lines). 

%%%%%%%%%%%%%%%%%%%%%%%%%%%%%%%%%%%%%%%%%%%%%%%%%%%%%%%%%%%%%%%
\subsection{Accessible scattering phenomena}

\begin{figure*}
    \centering
    \includegraphics{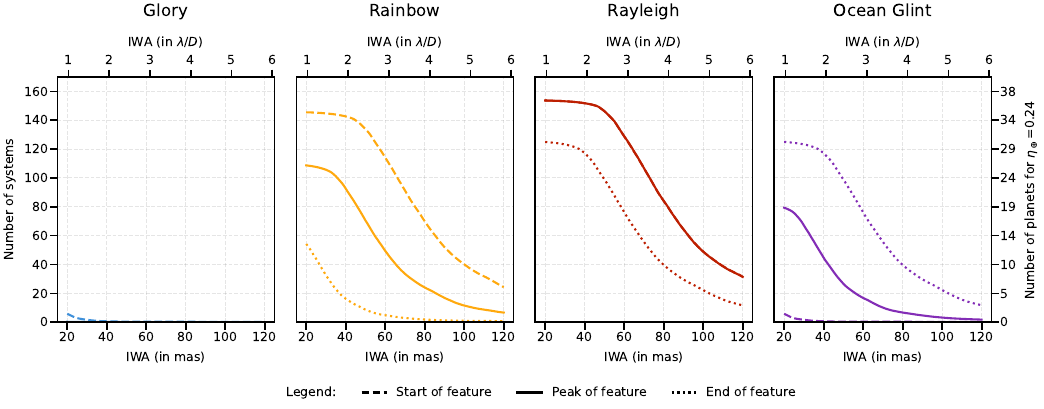}
    \script{create-figure-8-n-over-iwa.py}
    \caption{
        The number of systems for which key optical phenomena (Rayleigh scattering peak, rainbow, ocean glint, glory) would be accessible as functions of the \IWA.
        Each subplot contains three lines showing the number of systems where the start of each the phenomenon is accessible (dashed lines), where the peak is accessible (solid lines), and where the end is accessible (dotted lines).
        For Rayleigh scattering, the line for the start of the feature (at $\alpha=\qty{70}{\degree}$) and the line for the peak (at $\alpha=\qty{110}{\degree}$) coincide, as their respective phase angles are at the same distance from quadrature.
        The $x$-axis at the bottom shows the \IWA of the coronagraph in \unit{mas}, and the $x$-axis at the top shows the same \IWA converted to $\lambda/D$ for $\lambda = \qty{600}{\nano\meter}$ and $D = \qty{6}{\meter}$.
        The $y$-axis on the right indicates the number of systems where all phenomena would be accessible, assuming \qty{24}{\percent} of them have an Earth-like exoplanet in their HZ (as assumed in \Decadal).
    }
    \label{fig:accessible_phase_angles}
\end{figure*}

\Cref{fig:accessible_phase_angles} and Table \ref{tab:nstars_detect} show the number of systems in which key scattering phenomena will be accessible for different \IWA.
The Rayleigh scattering peak in polarization, which can be used to determine if a planet has an atmosphere, appears near quadrature and is therefore accessible in the most systems.
The rainbow feature appears further from quadrature (around $\alpha=\qty{40}{\degree}$), but will still be accessible in a significant fraction of systems.
The position and shape of the rainbow peak is sensitive to the properties of the cloud droplets, allowing their characterization if the peak of the rainbow feature can be reached.
In most systems, only the first half of the glint feature would be accessible, but this can still be used to identify a liquid ocean on the surface of a planet, a key indicator of habitability. 

%%%%%%%%%%%%%%%%%%%%%%%%%%%%%%%%%%%%%%%%%%%%%%%%%%%%%%%%%%
\begin{table}
    \centering
    \caption{
        The expected number of targets for which the peak phase angle of each phenomenon would be detectable when assuming randomly inclined, eccentric orbits (see \cref{sec:eccentric}).
        For reference, the \HWO target list consists of 164 targets in total.
    }
    \label{tab:nstars_detect}
    
% ATTENTION:
% THIS FILE IS GENERATED AUTOMATICALLY BY `create-figure-8-n-over-iwa.py`
% PLEASE DO NOT EDIT MANUALLY

% LAST UPDATE: 2023-07-27T14:23:34.453770Z 
% GENERATED ON: fv-az622-735 

\begin{tabular}{lrrrr}
\toprule
             & \multicolumn{4}{c}{Inner Working Angle (IWA)} \\
\cmidrule(lr){2-5}
 Feature     &   21 mas &   41 mas &   62 mas &   83 mas \\
\midrule
 Glory       &        0 &        0 &        0 &        0 \\
 Ocean Glint &       79 &       43 &       16 &        6 \\
 Rainbow     &      109 &       90 &       46 &       22 \\
 Rayleigh    &      154 &      152 &      125 &       78 \\
\bottomrule
\end{tabular}

    \script{create-figure-8-n-over-iwa.py}
\end{table}
%%%%%%%%%%%%%%%%%%%%%%%%%%%%%%%%%%%%%%%%%%%%%%%%%%%%%%%%%%

%%%%%%%%%%%%%%%%%%%%%%%%%%%%%%%%%%%%%%%%%%%%%%%%%%%%%%%%%%%%%%%
\subsection{Contrast curves}
\label{sec:results_contrast}

The contrast of the planet is also an important factor in assessing the detectability of these scattering phenomena. 
\Cref{fig:contrasts} shows approximate contrast curves for three of the stars in the \HWO target list, assuming an Earth-like planet on an edge-on orbit.
The overall contrast varies significantly for each system and throughout the orbit. 
In general, planets orbiting high-mass stars tend to offer larger angular separation but lower contrast ratios. 
The accessibility of these planets will depend on the starlight suppression of the coronagraph, which is influenced by its \IWA.
It will be important to consider both the accessible phase angles and contrast limits when deciding on the \IWA of the \HWO coronagraph if these scattering features are to be observed.
\looseness=-1

%%%%%%%%%%%%%%%%%%%%%%%%%%%%%%%%%%%%%%%%%%%%%%%%%%%%%%%%%%
% Figure 8
%%%%%%%%%%%%%%%%%%%%%%%%%%%%%%%%%%%%%%%%%%%%%%%%%%%%%%%%%%
%%%%%%%%%%%%%%%%%%%%%%%%%%%%%%%%%%%%%%%%%%%%%%%%%%%%%%%%%%

%%%%%%%%%%%%%%%%%%%%%%%%%%%%%%%%%%%%%%%%%%%%%%%%%%%%%%%%%%%%%%%%%%%%%%%%%%%%%%%%%%%%%%%%%%%%%%%%%%%%%%%%%%%%%%%%%%%%%%%%%%%%%%
\vspace{-2mm}
\section{Discussion}
\label{sec:4}
%%%%%%%%%%%%%%%%%%%%%%%%%%%%%%%%%%%%%%%%%%%%%%%%%%%%%%%%%%%%%%%

In this work, we consider each orbit to have a semi-major axis for which it would receive the same flux as Earth assuming a circular orbit.
As Earth is at the inner edge of the Habitable Zone, our estimates for the number of systems with accessible scattering features is conservative.
However, as we are only considering the accessibility of scattering features, it is possible to scale the results presented here for different semi-major axes. 
This is because the range of accessible features is determined only by the amount of the orbit covered by the mask, therefore increasing the semi-major axis of an orbit (e.g., double) is equivalent to decreasing the apparent mask size by the same factor (e.g., half).
\looseness=-1

%%%%%%%%%%%%%%%%%%%%%%%%%%%%%%%%%%%%%%%%%%%%%%%%%%%%%%%%%%%%%%%%%%%%%%%%%%%%%%%%%%%%%%%%%%%%%%%%%%%%%%%%%%%%%%%%%%%%%%%%%%%%%%
\vspace{-2mm}
\section{Conclusions}
\label{sec:5}
%%%%%%%%%%%%%%%%%%%%%%%%%%%%%%%%%%%%%%%%%%%%%%%%%%%%%%%%%%%%%%%

The Habitable Worlds Observatory is a coronagraphic space telescope envisioned to detect and characterize Earth-like exoplanets in the 2040s. 
The instrument design and capabilities are still under development and should be driven by the mission's science goals.
In this work, we investigated the possibility of using \HWO to detect optical phenomena with specific angular features in polarised and unpolarised reflected light, such as rainbows, ocean glint, and glories, which could establish a planet's habitability. 
We primarily focused on the range of accessible phase angles given the limitations of the coronagraphic \IWA and aimed to investigate which phenomena could potentially be observed.
While we do not consider the detectability of the scattering features beyond their accessibility, we show they are typically more distinct in polarised light, but the exoplanet-star contrast ratio is lower than the unpolarised light contrast ratio.

The desire to detect and characterize a large number of Earth-like exoplanets, and the intrinsic occurrence of such planets around different stellar spectral types, will drive the choice of coronagraphic \IWA for \HWO. 
Assuming a \qty{62}{\mas} \IWA, $\sim\qty{76}{\percent}$ of planetary systems in the NASA Exoplanet Exploration Program's Mission Star List for \HWO would be accessible at phases angles with maximal polarisation from Rayleigh scattering.
The rainbow of light that is scattered in droplets of water clouds and ocean glint, both of which indicate the presence of a water cycle which provides additional context to the habitability of an exoplanet, would be accessible in $\sim\qty{28}{\percent}$ and $\sim\qty{10}{\percent}$ of systems respectively.
The presence of a water cycle on Earth-like exoplanets in the remaining systems would have to be established through less direct methods: spectropolarimetry or rotational mapping.
To observe more scattering features in more systems, \HWO would need a smaller \IWA by use of a better-controlled coronagraph and potentially a larger primary mirror.

%%%%%%%%%%%%%%%%%%%%%%%%%%%%%%%%%%%%%%%%%%%%%%%%%%%%%%%%%%%%%%%%%%%%%%%%%%%

\vspace{-2mm}
\section*{Data Availability}

The NASA ExEP target list containing the data used for our simulations is  available online.%
\footnote{\url{https://exoplanets.nasa.gov/exep/science-overview/}}
This paper has been prepared using the \texttt{showyourwork!} framework \citep{Luger2021}, which makes the entire Python code to create all figures and the associated data available in an online repository to facilitate the reproducibility of our work.%
\footnote{\url{https://github.com/mkenworthy/HWObows/}}
\looseness=-1

%%%%%%%%%%%%%%%%%%%%%%%%%%%%%%%%%%%%%%%%%%%%%%%%%%%%%%%%%%%%%%%%%%%%%%%%%%%
\section*{Acknowledgements}

The 2023 \emph{Optimal Exoplanet Imagers} workshop, which sparked the work presented in this manuscript, was made possible thanks to the logistical and financial support of the Lorentz Center, Leiden, The Netherlands.
The research presented in this paper was initiated at a workshop held in Leiden in February 2023 and was partially supported by NOVA (the Netherlands Research School for Astronomy) and by the European Research Council (ERC) under the European Union's Horizon 2020 research and innovation programme (grant agreement \textnumero~866001---EXACT).

SRV and JLB acknowledge funding from the European Research Council (ERC) under the European Union's Horizon 2020 research and innovation program under grant agreement \textnumero~805445.
TDG acknowledges funding from the Max Planck ETH Center for Learning Systems.
SLC acknowledges support from an STFC Ernest Rutherford Fellowship. 
KB acknowledges support from NASA Habitable Worlds grant \textnumero~80NSSC20K152, and previous support for related work from NASA Astrobiology Institute's Virtual Planetary Laboratory under Cooperative Agreement \textnumero~NNA13AA93A.
IL acknowledges the support by a postdoctoral grant issued by the Centre National d'Études Spatiales (CNES) in France.
PB, IL, and YG were supported by the Action Spécifique Haute Résolution Angulaire (ASHRA) of CNRS/INSU co-funded by CNES.
EHP is supported by the NASA Hubble Fellowship grant \#HST-HF2-51467.001-A awarded by the Space Telescope Science Institute, which is operated by the Association of Universities for Research in Astronomy, Incorporated, under NASA contract NAS5-26555.
LA and EC acknowledge funding from the European Research Council (ERC) under the European Union's Horizon Europe research and innovation programme (ESCAPE, grant agreement \textnumero~101044152).
OA and LK acknowledge funding from the European Research Council (ERC) under the European Union's Horizon 2020 research and innovation programme (grant agreement \textnumero~819155).
SYH was funded by the generous support of the Heising-Simons Foundation.
RJP is supported by NASA under award number 80GSFC21M0002.
OHS acknowledges funding from the Direction Scientifique Générale de l'ONERA (ARE Alioth).

This work has made use of \textsf{numpy} \citep{NumPy2020}, \textsf{scipy} \citep{scipy_2020}, \textsf{matplotlib} \citep{matplotlib2007}, and \textsf{astropy}\footnote{\url{https://www.astropy.org}}, a community-developed core Python package and an ecosystem of tools and resources for astronomy \citep{astropy:2013, astropy:2018, astropy:2022}.
The colourblind-friendly colour schemes used in this paper were taken from \citet{Petroff_2021}.
This research has made use of NASA's Astrophysics Data System Bibliographic Services and the SIMBAD database, operated at CDS, Strasbourg, France. 

%%%%%%%%%%%%%%%%%%%%%%%%%%%%%%%%%%%%%%%%%%%%%%%%%%%%%%%%%%%%%%%%%%%%%%%%%%%
% References
\bibliographystyle{mnras}
\bibliography{bib}

%%%%%%%%%%%%%%%%%%%%%%%%%%%%%%%%%%%%%%%%%%%%%%%%%%%%%%%%%%%%%%%%%%%%%%%%%%%
% Author affiliations (load from file)
\section*{Author affiliations}
\begingroup
    \itshape
    \raggedright
    
% DO NOT EDIT THIS MANUALLY
% ALL ENTRIES ARE GENERATED BY A SCRIPT
% IF YOU CHANGE THINGS HERE IT WILL GET OVERWRITTEN

\affdest{1}Astrophysics, Department of Physics, University of Oxford, Denys Wilkinson Building, Keble Road, Oxford OX1 3RH, UK \\[0.5mm]
\affdest{2}Max Planck Institute for Intelligent Systems, Max-Planck-Ring 4, 72076 Tübingen, Germany \\[0.5mm]
\affdest{3}ETH Zurich, Institute for Particle Physics \& Astrophysics, Wolfgang-Pauli-Str. 27, 8092 Zurich, Switzerland \\[0.5mm]
\affdest{4}Department of Earth and Planetary Sciences, University of California, Riverside, CA 92521, USA \\[0.5mm]
\affdest{5}NASA Nexus for Exoplanet System Science, Virtual Planetary Laboratory Team, Seattle, WA 98195, USA \\[0.5mm]
\affdest{6}NASA Nexus for Exoplanet System Science, Terrestrial Polarization Team, Orlando, FL 32826, USA \\[0.5mm]
\affdest{7}Centre for Exoplanet Research, School of Physics and Astronomy, University of Leicester, University Road, Leicester, LE1 7RH, UK \\[0.5mm]
\affdest{8}Department of Earth \& Planetary Sciences and Department of Physics, McGill University, 3600 rue University, Montréal, QC, H3A 2T8, Canada \\[0.5mm]
\affdest{9}Leiden Observatory, Leiden University, P.O. Box 9513, 2300 RA Leiden, The Netherlands \\[0.5mm]
\affdest{10}SRON Netherlands Institute for Space Research, Niels Bohrweg 4, 2333 CA, Leiden, The Netherlands \\[0.5mm]
\affdest{11}LESIA, Observatoire de Paris, Université PSL, CNRS, Sorbonne Université, Université de Paris, 92195 Meudon, France \\[0.5mm]
\affdest{12}Department of Physics, University of California, Santa Barbara, CA 93106, USA \\[0.5mm]
\affdest{13}Department of Geoscience \& Remote Sensing, Delft University of Technology, Stevinweg 1, 2628 CN, Delft, The Netherlands \\[0.5mm]
\affdest{14}Royal Netherlands Meteorological Institute (KNMI), Utrechtseweg 297, 3731 GA, de Bilt, The Netherlands \\[0.5mm]
\affdest{15}Delft University of Technology, Kluyverweg 1, 2629 HS Delft, The Netherlands \\[0.5mm]
\affdest{16}STAR Institute, Université de Liège, Allée du six Août 19c, 4000 Liège, Belgium \\[0.5mm]
\affdest{17}Aix Marseille Université, CNRS, CNES, LAM, Marseille, France \\[0.5mm]
\affdest{18}NASA Ames Research Center, Bldg. 245, Moffett Field, USA \\[0.5mm]
\affdest{19}Université Grenoble Alpes, CNRS, IPAG, 38000 Grenoble, France \\[0.5mm]
\affdest{20}Department of Astronomy, California Institute of Technology, Pasadena, CA, 91125, USA \\[0.5mm]
\affdest{21}Department of Astronomy \& Astrophysics, University of California, Santa Cruz, CA 95064, USA \\[0.5mm]
\affdest{22}Steward Observatory, University of Arizona, 933 North Cherry Avenue, Tucson, AZ 85719, USA \\[0.5mm]
\affdest{23}DTIS, ONERA, Université Paris Saclay, 91123 Palaiseau, France \\[0.5mm]
\affdest{24}DOTA, ONERA, 92322 Châtillon, France \\[0.5mm]
\affdest{25}Subaru Telescope, NAOJ, USA \\[0.5mm]
\affdest{26}College of Optical Sciences, University of Arizona, Tucson, AZ 87521, USA \\[0.5mm]
\affdest{27}Astrobiology Center, 2 Chome-21-1, Osawa, Mitaka, Tokyo, 181-8588, Japan \\[0.5mm]
\affdest{28}NASA Goddard Space Flight Center, 8800 Greenbelt Rd, Greenbelt, MD  20771, USA \\[0.5mm]
\affdest{29}University of Maryland Baltimore County, 1000 Hilltop Cir, Baltimore, MD 21250, USA \\[0.5mm]
\affdest{30}Faculty of Aerospace Engineering, TU Delft, Building 62, Kluyverweg 1, 2629 HS Delft, The Netherlands \\[0.5mm]
\affdest{31}National Research Council Canada, Herzberg Astronomy and Astrophysics Research Center, Victoria, B.C., Canada \\[0.5mm]
\affdest{32}European Space Agency, ESTEC, Keplerlaan 1, 2200 AG Noordwijk, The Netherlands \\[0.5mm]
\affdest{33}Université Côte d'Azur, Observatoire de la Côte d'Azur, CNRS, Laboratoire Lagrange, France \\[0.5mm]
\affdest{34}Space Telescope Science Institute, 3700 San Martin Drive, Baltimore, MD 21218, USA \\[0.5mm]

\endgroup

\end{document}